# Time-Resolved Photoemission Spectroscopy of Quantum Materials Using High Harmonic Generation: Probing Electron-Phonon Interactions and Non-Equilibrium Dynamics

Takeshi Suzuki[1,*] and Kozo Okazaki[1,2,3,*]

[1]*Institute for Solid State Physics, The University of Tokyo, Kashiwa, Japan*

[2]*Trans-scale Quantum Science Institute, The University of Tokyo, Tokyo, Japan*

[3]*Material Innovation Research Center, The University of Tokyo, Kashiwa, Japan*

*Corresponding authors: takeshi.suzuki@issp.u-tokyo.ac.jp; okazaki@issp.u-tokyo.ac.jp

**Abstract**

Recent advancements in ultrafast laser systems and high harmonic generation (HHG) techniques have enabled time-resolved photoemission spectroscopy on femtosecond timescales, opening up unprecedented opportunities to explore quantum materials in both time and momentum space. In this review, we present recent representative studies utilizing HHG-laser-based time- and angle resolved photoemission spectroscopy for a variety of quantum materials. We particularly highlight electron-phonon interactions and non-equilibrium dynamics in time and frequency domain, through which rich information about non-equilibrium electron-phonon couplings and related phenomena has been clearly revealed.

# 1 Introduction

Ultrafast phenomena have enabled detailed investigations of the fundamental interactions that govern the behavior of matter across various scientific disciplines. Initially, the concept of ultrafast science emerged from physics, where it provided an unprecedented ability to probe atomic and molecular dynamics [1]. However, the impact of ultrafast science extends far beyond physics into chemistry, biology, and materials science, enabling breakthroughs in diverse areas such as reaction dynamics [2], molecular transformations [3], and studies of biological processes at a molecular level [4]. By offering insights into the dynamics of systems that were previously inaccessible by conventional techniques, ultrafast science has become an indispensable tool across multiple research domains.

In condensed matter physics, ultrafast techniques have provided a powerful approach of probing the nonequilibrium dynamics of quantum materials, where electron-electron and electron-phonon interactions drive the system far from equilibrium [5]. These interactions, often occurring on timescales much shorter than traditional measurement techniques can resolve, are fundamental to understanding the properties of materials, particularly in systems exhibiting emergent phenomena such as superconductivity [6], magnetism [7], and topological behavior [8]. As a result, ultrafast probes have emerged as critical tools for studying quantum materials, enabling us to directly observe how electronic and lattice degrees of freedom interact and evolve in real-time.

Among the techniques that can capture the electronic properties in nonequilibrium state of quantum materials, time-resolved photoluminescence has often been used in the first stage of investigating condensed matter systems to gain insights into carrier recombination dynamics and the behavior of excitons in semiconductors [9]. However, the time-resolution is ultimately limited by the recombination rate or the lifetime of photo-excited carriers, which is usually in the order of picoseconds to nanoseconds. Another powerful technique often employed to detect the ultrafast phenomena is time-resolved reflectivity or absorption, where the ultrafast short-pulsed laser is split to pump and probe beams, then pump pulse excites the material and after some delay time probe beam detects the reaction of pump induced dynamics [10]. While the time resolution of time-resolved reflectivity or absorption is limited by the laser pulse duration, which is typically on the order tens of femtoseconds, the information obtained is not always straightforward to interpret, as these techniques probe changes in the dielectric function at the wavelength of the probe pulse, usually in the near-infrared to visible range.

Time- and angle-resolved photoemission spectroscopy (TARPES) has emerged as a very powerful tool because it can directly capture the dynamic evolution of the electronic structure on femtosecond timescale. Fig. 1(a) shows the schematics of TARPES measurements. Direct access to electronic properties with momentum resolution is a distinctive advantage, enabling the investigation of changes in the electronic band structure and carrier dynamics in momentum space. The time resolution, being solely governed by the pump and probe pulse durations, enables the observation of ultrafast phenomena such as carrier dynamics, the formation of excitons (Fig. 1(b)) [11] [12] [13], non-Fermi-Dirac distributions [14] [15], and photo-induced phase transitions (Fig. 1(c)) [16] [17] [18]. Furthermore, the creation of the light-dressed electronic states, such as the Floquet-Bloch state, is directly ovserved, providing all the information needed to prepare for Floquet engineering (Fig. 1(d)) [19] [20] [21].

In the early stages of TARPES, a photon energy of 6 eV was commonly used for the probe, and up-conversion techniques with nonlinear crystals such as β-$BaB_2O_4$(BBO) were employed [22] [23]. Recently, 7 eV has been also used for the probe using $KBe_2BO_3F_2$ optical crystals [24]. While a 6 eV light source is convenient to use: it provides sufficient photon flux with relatively low fundamental light fluence and can pass through the air, the accessible momentum range is limited, and it can only detect the band structures near the Γ point. High harmonic generation (HHG) based on rare gases has emerged as an alternative platform for use as a probe [25] [26] [27] [28] [29] [30] [31] [32] [33] [34] [35] [36] [37]. The photon energy can be extended into the extreme ultraviolet (EUV) region ($h\nu > 10$ eV), expanding the accessible momentum range to cover the 1st Brillouin zone for most materials. Further extension into the soft X-ray region ($h\nu > 100$ eV) allows for the investigation of the core-level dynamics, enabling element-specific studies [38].

In this review, we focus on HHG-laser-based TARPES (HHG laser TARPES) and highlight its applications in studying the nonequilibrium dynamics in quantum materials. Particular emphasis is placed on electron-phonon coupling, which plays a crucial role in carrier dynamics, photo-induced phase transitions, and dynamical collective modes [39]. In section II, we describe the technical aspects of HHG laser TARPES, introducing our experimental setup. Section III presents TARPES studies on representative quantum materials, including monolayer and bilayer graphene, iron-based superconductors, excitonic insulators, and charge density wave (CDW) materials. Finally, in section IV, we discuss future prospects on HHG laser TARPES from both experimental and theoretical perspectives.

## 2 Experiment

Figure 2 shows the configuration of our TARPES setup [40]. Two Ti:sapphire amplifiers are employed, each with a center wavelength of 800 nm, a pulse duration of 35 fs, a pulse energy of 0.7 mJ, and a repetition rate of 10 kHz. Compared to 1 kHz, space charge effects are much less evident, which enables us to obtain sufficient signal. In this setup, one amplifier serves as the pump source, while the other is used for the probe. Both amplifiers are seeded by a common Ti:sapphire oscillator, which significantly reduces timing jitter between the two, thereby preserving that the time resolution of the system is not degraded. This configuration provides a notable advantage over typical setups that rely on electrical synchronization techniques. Additionally, using twin amplifiers enables a total output power exceeding 10 W, which would typically require cryogenic cooling of the gain crystal in a single-amplifier setup to avoid thermal loading.

For the probe, the photon energy is first doubled to 3.10 eV by generating a second harmonic (SH) of the fundamental light using a BBO crystal. This pulse is then focused into an argon gas-filled cell to induce HHG. The gas is approximately 4 mm thick, with the argon back pressure maintained between 3 and 5 Torr. Since the HHG output is highly sensitive to the focused beam position, a beam stabilizer is employed. This system steers two mirrors to form a feedback loop, effectively suppressing both thermal drifts and fast mechanical fluctuations. The seventh harmonic, corresponding to 21.7 eV, is typically selected using a pair of SiC/Mg multilayer mirrors (MLM). Differential pumping between the HHG and MLM chambers allows a relatively high pressure ($\sim 10^{-3}$ Torr) in the HHG chamber, while maintaining high vauum ($\sim 10^{-8}$ Torr) in the MLM chamber. To block the fundamental light, 50-100 nm thick aluminum or zirconium filters are used. The total photon flux at the sample position is estimated to be $\sim 10^9$ photons/s.

For sample characterization, a He discharge lamp is integrated into the system to measure photoemission spectra under equilibrium conditions. The pump source is an optical parametric amplifier (OPA), providing wavelength-tunable excitation between 1160 and 2580 nm. The temporal profile of the pump is monitored using an autocorrelator. A delay stage is inserted into the pump beam path to adjust the time delay between the pump and the probe pulses. Half-wave ($\lambda/2$) and quarter-wave ($\lambda/4$) plates are utilized to control the polarization of the pump, while a variable attenuator is employed to adjust the pump fluence. Additionally, the system can switch to the fundamental wavelength of 800 nm by activating a flipping mirror.

Photoelectrons are detected using a hemispherical electron analyzer, with which the energy resolution can be improved better than 250 meV by changing the slit size and pass energy, with the analyzer chamber maintained at a base pressure of approximately $2 \times 10^{-11}$ Torr. The cryostat can cool the sample to a temperature as low as 7 K.

3   Nonequilibrium phenomena in Quantum Materials
   3.1   Graphene

   Graphene, a carbon-based material, has attracted significant attention owing to its unique physical, electronic, and chemical properties, prompting extensive studies. Among these, its optical properties have attracted particular interest, with several remarkable phenomena reported, including multiple carrier generations and phonon bottleneck effects. These phenomena originate from the dynamics of fermions within a linearly dispersed band structure, known as the Dirac cone [41]. The introduction of a second graphene layer, referred to as bilayer graphene, modifies the massless band structure of monolayer graphene, and induces a transition to a massive band structure. Consequently, the carrier dynamics in bilayer graphene exhibit behaviors distinct from those observed in monolayer graphene. Moreover, varying the twisting angle between the two layers in bilayer graphene can lead to substantial modifications in its electronic properties. In this context, TARPES is an exceptionally suitable technique, as it enables direct observation of the ultrafast evolution of electronic states in the band structure following photoexcitation. Furthermore, the relatively high photon energies provided by HHG are crucial for accessing the Dirac cone, which is located at the edge of the Brillouin zone. We utilized HHG laser TARPES to investigate the carrier dynamics in a monolayer graphene and quasi-crystalline twisted bilayer graphene. Our findings demonstrate that the carrier dynamics are highly sensitive to the structural characteristics of the layers.

   3.1.1 Monolayer graphene

   Monolayer graphene is a "supermaterial" known for its exceptional strength, lightness, and ultrahigh electrical and thermal conductivity [41] [42] [43]. Its optical properties have also attracted attention for use in optoelectronics and nanophotonics. However, the mechanisms of photo-induced phenomena in graphene, such as multiple carrier generation [44]

[45] [46] [47] [48], remain not fully understood, highlighting the need for further investigation into non-equilibrium carrier dynamics. In the energy dissipation process of hot carriers, optical phonons typically play a crucial role through carrier-phonon scattering. This is because optical phonons possess high energies and relatively weak momentum dispersion, which enables more efficient scatterings channels compared to acoustic phonons. A new cooling mechanism, referred to as "supercollisions" (SCs) [49], has been proposed. In this process, hot carriers lose their energy through a three-body scattering involving an electron, a defect, and an acoustic phonon. While scattering by acoustic phonons typically transfers less energy compared to optical phonons, disorder-mediated interactions enable acoustic phonons to access a wider phase space, making SCs an efficient energy dissipation pathway.

We first investigated the hot carrier distribution in *n*-type epitaxial Graphene on SiC [50]. Figures 3(a) shows the temporal evolution of the electron temperature (left axis) and the peak intensity of the energy distribution curves (EDCs) derivative (right axis), measured with an excitation photon energy of 3.1 eV and a fluence of 0.32 mJ/cm$^2$. The electronic temperature was determined by fitting the angle-integrated spectra with the function

$$I(E,t) = \int D(E')f(E',T_e,\mu)G(E-E',\sigma)dE', \quad (1)$$

where $I(E,t)$ is the photoemission intensities at a delay time $t$, $D(E)$ is the density of states of graphene, $G(E,\sigma)$ is a Gaussian with a width, $\sigma$, corresponding to the experimental energy resolution, and $f(E',T_e,\mu)$ is the Fermi-Dirac distribution function at the electron temperature $T_e$ [51] [52] [53] and the chemical potential $\mu$. The electronic temperature rises as high as 900 K and then relaxes with the time constant of ~ 400 fs. This value is similar to the previous report for *p*-type quai-free-standing graphene on the H-terminated SiC substrate [54] [55]. The relaxation mechanism is attributed to electron-electron scattering where the Coulomb collisions take place between electrons within the same valley (*intra-valley scattering*).

From the obtained electron temperature, the chemical potential shift relative to the Fermi level ($E_F$) can be estimated as shown in Fig. 3(b) by the yellow circles. In equilibrium ARPES, the chemical potential of the sample is equilibrated with the grounded substrate through charge transfer.

The observed shift of the chemical potential after photoexcitation thus suggests that such charge compensation did not occur within the picosecond timescale. The hot electron distribution was previously reported to accompany cascade carrier multiplication via the Auger process [47], and our observation is also consistent with this process.

We next studied the graphene grown on a SiC(0001) C-terminated surface, which exhibits the high intrinsic carrier mobility (over 100,000 cm²/V·s) [56]. This high mobility allowed us to suppress the SC cooling process, enabling the investigation of the intrinsic relaxation dynamics of photoexcited carriers in graphene via electron-optical-phonon scattering and optical-phonon-acoustic-phonon decay processes. Figures 3(c) and 3(d) present TARPES images at delay time t = -1.0 ps, and 0.1 ps. To highlight the pump-induced change, the differential image is also displayed in Fig. 3(e). The excitation fluence is 0.66 mJ/cm² and the pump photon energy is 1.55 eV. While only the lower Dirac cone is visible before the irradiation, the upper Dirac cone appears at the originally unoccupied region after the irradiation. From the fitting analysis by the equation (1) for angle-integrated spectra, the time-dependent electron temperature is obtained as shown in Fig. 3(f).

In order to gain quantitative insight into the energy dissipation of the electron system via multiple electron-phonon scatterings, we performed theoretical calculations based on the two-temperature model [51] [53]. Time-dependent electronic and optical-phonon temperatures are shown as red-solid and black-dashed lines, respectively, in Fig. 3(f). Figure 3(g) shows the term-by-term comparisons of the calculation results for $dT_e/dt$; specifically, the heating/cooling rates originating from a pumping laser, optical phonons, and SCs are displayed separately. By comparing the energy dissipation via SCs (integral of the blue area) with the total energy dissipation (sum of integrals of the yellow and blue areas), we find that SCs account for only 1.1±0.1 % of the total energy loss. This indicates that SCs have a negligible effect on the reduction of electronic temperature in extremely high-mobility graphene. This finding in C-faced graphene is in contrast to the case of Si-faced graphene, where SCs are more frequent and play a more dominant role in the cooling process [54].

3.1.2 Quai-crystalline twisted bilayer graphene

Twisted bilayer graphene (TBG) exhibits a range of exotic quantum phenomena, with the twist angle (Θ) serving as a key factor in exploring various states of matter [57] [58] [59] [60] [61], such as two-dimensional superconductivity (Θ = 1.1°) [62] [63] and a two-dimensional quasicrystal (Θ = 30°) [64] [65] as shown in Fig. 4(a). In QCTBG, the interlayer interaction is minimal due to the 30° twist angle, and the Dirac Fermions in both layers remain massless. Figure 4(b) shows a schematic illustration of the electronic band structure of QCTBG. The outer red and blue Dirac cones represent the upper layer Dirac (ULD) and lower layer Dirac (LLD) cones, respectively. As a result of the strong interlayer interaction connected by the Umklapp scattering in each layer, the replica bands of the ULD and LLD are formed, as shown in the inner red and blue Dirac cones in Fig. 4(b). This structure increases the two-dimensional density of Dirac Fermions, influencing the carrier dynamics. Furthermore, the epitaxial graphene layer, as studied here, is complicated with various factors such as interface, substrate, and doping, compared with a free-standing layer of non-doped graphene.

We performed the TARPES measurements for QCTBG and investigated how the electronic temperatures and occupations are modified in each layer after the optical excitation [66]. The angle-integrated spectra were fitted using equation (1) at each delay time. Figure 4(c) shows the temporal chemical potential shifts for the ULD, LLD, and NTBG. Surprisingly, the opposite behavior was observed between the ULD and LLD, that is, ULD underwent a negative shift, whereas LLD underwent a positive shift. It was also found that the chemical potential shift for the NTBG band remained constant at essentially zero over the delay time from 0.1 to 0.6 ps. The striking difference among the three types of Dirac cones provides clear evidence of a carrier imbalance between the ULD and LLD bands in QCTBG at the ultrafast time scale.

To investigate the underlying mechanism of the observed carrier imbalance, we employed a set of rate equations. The model used in this study is schematically shown in Fig. 4, which considers carrier transport between the ULD and LLD, indicated by red ($\gamma_1$) and blue ($\gamma_2$) arrows. We also take into account photo-induced carrier injection from the buffer layer into the ULD ($G_1$) and LLD ($G_2$). Our analysis revealed that $\gamma_1 > \gamma_2$ indicates more frequent carrier transfer between the graphene layers than between the LLD and the substrate. Furthermore, the finite values of $G_1$ and $G_2$ demonstrate the presence of transient carrier doping from the substrate into each graphene layer. A larger $G_2$ compared to $G_1$ is identified as the origin of the observed unbalanced carrier imbalance between the ULD and LLD.

We also investigated the environmental effects of the carrier dynamics of QCTBG [67]. We first measured two types of QCTBG; one with equivalent chemical potentials between the ULD and LLD, and the other with inequivalent ones. Both types of QCTBG exhibited larger values of $\gamma_1$ than $\gamma_2$, indicating that the carrier transport between the graphite layers (UL and LL) is faster than that between the graphene and the SiC substrate. Furthermore, both types showed $G_2 > G_1$, with a ratio of $G_1/G_2 \approx 0.4$-$0.6$. This ratio is consistent with a carrier transport model involving exponential decay with distance. The model assumes that the $G_i \propto \exp(-2\kappa d_i)$, where $d_i$ is the distance to the graphene layer shown in Fig. 4(e) and $\kappa$ is the decay constant of the electronic state that donates carriers to graphene, estimated to be ~ 0.074-0.13 Å$^{-1}$. To further examine carrier transport between graphene and the SiC substrate, we performed TARPES measurements on single-layer graphene grown on a stepped SiC substrate, as shown in Fig. 4(f), and compared the results with those for a flat SiC substrate [50]. Our findings suggest that the source of doping carriers likely originates from interface step states.

3.2 Iron-based high-temperature superconductors

Iron-based superconductors are a class of high-temperature superconducting materials that have garnered significant attention in recent years due to their unique properties and potential applications. Discovered in 2008 [68], these materials, which include iron combined with elements such as arsenic, selenium, or phosphorus, exhibit superconductivity at second highest temperatures at ambient pressure, following cuprate superconductors [69]. In addition to superconductivity [70] [71] [72] [73], iron-based superconductors exhibit a range of intriguing phenomena, including electronic nematicity [74] [75] [76] [77], a Bardeen-Cooper-Schriefer (BCS) to Bose-Einstein condensation (BEC) crossover [78] [79] [80], and the emergence of Majorana fermions on topological superconducting surfaces [81] [82]. A number of time-resolved studies have been conducted on iron-based superconductors to uncover physical properties that remain hidden under equilibrium conditions. These include investigations into the melting dynamics of spin-density wave [83], nematic-orbital excitations [84], structural dynamics [85] [86] [87], temporally periodic formation of spin-density-waves [88], precise estimation of the deformation potential [89], and shifts in chemical potential [90]. Many of these phenomena are accompanied by the pronounced appearance of coherent phonons, which strongly couple to the electronic system and significantly influence its properties. We have also studied iron-based superconductors, specifically $BaFe_2As_2$ [91] and FeSe [92] using TARPES measurements. In both materials, we observed pronounced oscillations attributed to the generation of coherent phonons, and we discuss their relationship to superconductivity in detail.

### 3.2.1 $BaFe_2As_2$

One of the most prominent features of $BaFe_2As_2$ (Ba122), the parent compound of a iron-based high-temperature superconductor system, is the emergence of superconductivity under various conditions: hole doping via substitution of K for Ba [93], electron doping via substitution of Co for Fe [94], isovalent substitution of P for As [95], as well as under high pressure [96]. In the case of isovalent substitution and high-pressure application, superconductivity can be induced through structural modifications without carrier doping. This observation suggests the possibility that Ba122 could be driven into a superconducting state via photo-irradiation. Therefore, investigating the transient electronic structure following photoexcitation is essential to gain insights into the photoinduced superconducting phase.

Figure 5(a) shows the Fermi surface (FS) mapping of Ba122. Hole and electron FSs are located at the Z and X points, respectively. Figures 5(b) and 5(c) show the momentum-integrated TARPES spectra across the hole and electron FSs as a function of energy and pump-probe delay time. Both FSs exhibit an immediate increase in electronic population followed by a gradual relaxation, with oscillatory components superimposed on the background dynamics. These oscillations are more clearly visualized in Figs. 5(d) and 5(e), which show the oscillatory parts of the integrated intensities above $E_F$ extracted from the boxed regions in Figs. 5(b) and 5(c), respectively, along with cosine-function fits. The hole and electron FSs show anti-phase oscillations at 5.5 THz, corresponding to the $A_{1g}$ phonon mode (Fig. 5(f)). These cosine-like oscillations are characteristic of displacive excitation of coherent phonons (DECPs) [97], where photoexcitation modifies the adiabatic potential energy surface, leading to atomic displacements. Band structure calculations based on density functional theory (DFT) with modulated $A_{1g}$ phonon displacements revealed that the hole FS mainly originates from the $d_{z^2}$ orbital, which is strongly warped for lower pnictogen heights. Conversely, the electron FSs behave oppositely, becoming larger for higher pnictogen heights and smaller for lower ones. This inversion explains the observed anti-phase oscillations. in the lowering of the pnictogen height, as induced by P substitution for As, leads to superconductivity via structural modification without carrier doping [95].

### 3.2.2 FeSe

Unlike Ba122 and other iron-based superconductors, FeSe does not exhibit antiferromagnetic order, making it particularly suitable for studying the nature of superconductivity without the influence of magnetic ordering. Notably, FeSe shows a significant enhancement of its superconducting critical temperature ($T_c$) under various external modifications. Although $T_c$ is approximately 10 K at ambient pressure [98], it increases to around 40 K under applied pressure [99] [100] [101]. Furthermore, intercalation of spacer layers can raise $T_c$ to ~ 40 K [102] [103], and single-layer FeSe exhibits a $T_c$ of ~ 60 K [104]. The key ingredient for achieving higher $T_c$ values lies in engineering of the band structure. In this context, photoexcitation offers a promising approach for its manipulation.

We performed TARPES measurements for FeSe in a similar manner to Ba122. Fig. 6(a) shows the FS mapping of FeSe, where the hole and electron FSs are located at the Γ and M points, respectively. Figures 6(b) and 6(c) show the momentum-integrated TARPES spectra across the hole and electron FSs as a function of energy and pump-probe delay time. The oscillations extracted from these spectra are shown in Figs. 6(d) and 6(e), respectively. In contrast to Ba122, the oscillations for both the hole and electron FSs are in-phase, corresponding to the coherent phonons illustrated in Fig. 6(f). We further investigated the behavior at longer delay times and analyzed the leading-edge midpoint (LEM) shifts for the hole and electron FSs, shown as blue and red markers in Fig. 6(g), respectively, as a function of pump fluence. Based on carrier conservation, photoexcited electrons from the hole bands are expected to relax into the electron bands. However, the LEM shift of the electron FS decreases with increasing pump fluence, as shown in Fig. 6(g). This unexpected behavior can be understood by considering the overall LEM shifts, denoted as <Δ> in Fig. 6(g), with one possible origin being the presence of a superconducting order.

### 3.3 Excitonic Insulator

When light is absorbed in a semiconductor, an electron and hole pair is created, forming a bound state called "exciton" due to the attractive Coulomb interaction [105]. An exciton is charge-neutral and bosonic quasi-particle. Because photoexcited states in semiconductors are directly related to optoelectronic devices such as solar cells, optical switches, and semiconductor lasers, exciton physics has been extensively studied for a long time [105]. One of the most intriguing challenges is the realization of macroscopic quantum condensed states such as exciton BECs [106] [107] or electron-hole BCS states [108], and intensive research is still ongoing. On the other hand, when the exciton binding energy exceeds the band gap in a semiconductor, or the band overlap in a semimetal, a substantial number of excitons can be spontaneously created without excitation by light. This concept is known as an "excitonic insulator", whose phase diagram is shown in Fig. 7. Although the excitonic insulator was theoretically proposed more than half a century ago [109] [110] [111] [112], experimental evidence remained elusive for a long time. Recently, several candidate materials have been proposed, accompanied by characteristic signatures consistent with excitonic insulators. In this section, we highlight two representative materials, namely $1T$-TiSe$_2$ and Ta$_2$NiSe$_5$, for which TARPES measurements have played a crucial role in understanding their electronic states.

#### 3.3.1 $1T$-TiSe$_2$

$1T$-TiSe$_2$ is an indirect semimetal, with a hole band at the Γ point and an electron band at the L point. Below the CDW phase transition temperature of 200 K, the electron band is folded to the Γ point, leading to hybridization between the hole and electron bands. This raises a question: what is the driving force for the CDW formation? One of the most intriguing possibilities is excitonic instability. Equilibrium measurements, such as resistivity [113] and ARPES [114] [115], have reported characteristic behaviors consistent with those expected in an excitonic insulator. Figures 8(a) and 8(b) show ARPES images of $1T$-TiSe$_2$ above and below the CDW phase transition temperature [115]. The folded band dispersion observed below the transition can be reproduced by calculations that take excitonic effect into account. However, alternative mechanisms such as the simple nesting model [116] and the band-type Jahn-Teller model [117] [118] have also been proposed. Thus, the exact origin of the CDW in $1T$-TiSe$_2$ remains a subject of ongoing debate.

TARPES measurements have provided new insights into the physics of $1T$-TiSe$_2$ [119]. The basic idea is to measure the response time of the insulating state following intense photoexcitation. From this response time, the underlying insulating mechanism can be inferred [120]. Figure 8(c) shows a static ARPES image at 125 K, where the folded Se 4p hole band is clearly observed at the $\bar{M}$ point. Figure 8(d) shows a series of TARPES images at increasing pump-probe delay times. Following excitation, an electron-like band emerges, crossing the Fermi level and extending from the $\bar{M}$ to $\bar{\Gamma}$ points. Notably, the downward-dispersing folded Se 4p band at the $\bar{M}$ point disappears or is at least considerably reduced. Figure 8(e) shows the temporal evolution of the integrated intensity of the folded Se 4p band at various pump fluences. The drop time of the intensity becomes faster with increasing fluence. This fluence dependence is summarized in Fig. 8(f), which shows that the drop time $\tau$ follows a relation of $\tau \propto 1/\sqrt{n}$. This is characteristic of Coulomb interaction screening [121], supporting the excitonic origin as the driving mechanism of the CDW transition.

### 3.3.2 Ta$_2$NiSe$_5$

Ta$_2$NiSe$_5$ is a direct-gap semiconductor and undergoes a structural phase transition from orthorhombic to monoclinic structures below 328 K [122]. Different from $1T$-TiSe$_2$, there is no CDW phase in Ta$_2$NiSe$_5$, which makes it an ideal system for the quest of an excitonic insulator, and considerable attention has been paid to this material. Among many equilibrium studies [123], temperature-dependent ARPES shows characteristic band dispersions consistent with an excitonic insulator [124] [125] [126] [127]. Figure 9(a) shows the second derivative of the ARPES spectra as a function of temperature [125]. With decreasing temperature, the top of the valence band becomes flatter, which is a characteristic behavior of an excitonic insulator. Figure 9(b) shows the theoretical calculation of the single-particle excitation spectra for 40, 270, and 340 K. The flat-top valence band at 40 K is well reproduced and as attributed to the excitonic coupling between the valence and conduction bands.

To search for the unique profile hidden under the equilibrium conditions, many time-resolved studies have been conducted on $Ta_2NiSe_5$. Figure 9(c) shows time-resolved optical reflectivity at different temperatures [128]. Significant oscillations at multiple frequencies of 1, 2, and 3 THz are observed, and they are attributed to coherent phonon modes. Figure 9(d) shows the amplitude of the 1 THz mode as a function of temperature. The characteristic BCS-like behavior suggests that this mode is not a simple coherent phonon, but more likely a phonon-coupled collective mode [129] [130]. Figure 9(e) shows the TARPES measurements for $Ta_2NiSe_5$ [131]. Figure 9(f) shows time dependence of the energy position of the valence band top under different photon fluences. In the low-fluence regime below 0.2 mJ/cm$^2$, the band gap becomes narrower. In the high-fluence regime above 0.2 mJ/cm$^2$, on the other hand, the band gap increases, successfully demonstrating the ability to use light to control the material properties.

We also studied $Ta_2NiSe_5$ using TARPES [132]. Figures 10(a) and 10(b) display the energy-momentum (*E-k*) intensity maps around the Γ point at 100 K, before and after the pump pulse. To examine the collapse of the flat band, a feature associated with excitonic insulators, we show the temporal evolution of the integrated TARPES intensity in Fig. 10(c) for different pump fluences. The rectangular region in Fig. 10(a) indicates the integration range. The decreasing behavior of intensity strongly depends on the pump fluence and accelerates with increasing fluence. The drop time of the flat band ($\tau_{Flat}$) is plotted in Fig. 10(d). The gap collapse time in excitonic insulators is expected to be inversely proportional to the plasma frequency, and proportional to $1/\sqrt{n}$ [119] [121]. We found that the drop time was proportional to $F^{-0.7}$, where *F* is the pump fluence. The deviation from -0.5 is due to nonlinear absorption processes, such as two- or three-photon absorption. This behavior is similar to that of 1*T*-TiSe$_2$

[119], suggesting that Ta$_2$NiSe$_5$ is also an excitonic insulator. Figure 10(b) shows the time-integrated TARPES image after pumping, where both electron and hole bands cross $E_F$ at $k_F \sim \pm 0.1$ Å$^{-1}$. Under equilibrium conditions, hybridized Ni 3$d$ and Se 4$p$ orbitals mainly compose the single valence band near $E_F$, whereas Ta 5$d$ orbitals primarily form the doubly degenerate conduction bands. Therefore, the emergence of the electron and hole bands with the same $k_F$ indicates that the hybridization between the two Ta chains is sufficiently strong to lift the degeneracy. This result, which is not predicted by band structure calculations, reveals a surprising non-equilibrium metallic phase distinct from the equilibrium high-temperature phase.

To further investigate the photo-induced insulator-to-metal transition in Ta$_2$NiSe$_5$ through electron-phonon interactions, we developed a new analysis method called frequency-domain ARPES (FDARPES) [133], which was also used to investigate mode-selective electron-phonon couplings in $T$d-WTe$_2$ [134]. Figure 11(a) shows the differential TARPES images before and after pump excitation, with peak positions indicated by circles. To analyze the photo-induced changes, we focused on the electron-phonon couplings by examining the time-dependent intensities in regions I–IV shown in Fig. 11(b). The data include both background carrier dynamics and oscillations arising from coherent phonons. We extracted the oscillatory components by fitting the carrier dynamics with a double-exponential function as indicated by black solid lines in Fig. 11(b) and subtracting the fit, followed by Fourier transforms. The result, shown in Fig. 11(c), reveals distinct peak structures for each $E$-$k$ region. We then created FDARPES maps of the peak intensities for frequencies of 1, 2, and 3 THz (Figs. 11(d)-(f)). Each frequency corresponds to a specific coherent phonon mode. The FDARPES map for 2 THz exhibits the strongest signal around $E_F$, indicating that the corresponding coherent phonon mode, illustrated in Fig. 11(g), is most relevant to the photo-induced insulator-to-metal transition. This is consistent with a recent theory suggesting that FDARPES intensity is proportional to the electron-phonon coupling matrix elements [135]. By comparing the FDARPES maps with the band dispersions from TARPES, we found that the 1-THz map better matches the post-excitation dispersions, while the 3-THz map is closer to pre-excitation dispersions. This suggests that the 1-THz mode couples to semimetallic bands, while the 3-THz mode couples to semiconducting bands. These findings demonstrate that FDARPES can selectively detect electron-phonon coupling to different electronic states in a frequency-resolved manner.

3.4  CDW materials

CDW materials are a class of condensed matter systems where the electron density spontaneously modulates in a periodic fashion, typically due to electron-phonon interactions [136]. This phenomenon leads to a distortion of the crystal lattice, often accompanied by a transition from a metallic to an insulating state, and is commonly observed in low-dimensional materials such as transition metal dichalcogenides (TMDs) [137] and certain organic conductors [138]. While CDWs were originally explained by electron-phonon interactions [136], CDWs are also a striking manifestation of electron correlation effects and are believed to play a key role in various electronic properties, including superconductivity [139] [140] [141], magnetism [142] [143], and non-Fermi liquid behavior [144]. Over the past few decades, CDWs have attracted significant attention due to their rich phase diagrams, tunability via external parameters such as pressure and temperature [145], and their potential applications in novel electronic devices [146].

Photoexcitation has also served as a very powerful tool to perturb systems out of equilibrium and observe their responses. Especially, collective modes such as CDW amplitude or phase modes have been central topics and intensively studied for the insight into electron-electron and electron-phonon interactions [147] [148] [149]. Moreover, with increasing excitation fluence, some CDW materials exhibit photoinduced phase transitions, which have attracted significant interest for the potential application to the ultrafast switching [150] [151] [152] [153] [154] [155] [156] [157]. In the study of CDW dynamics, TARPES has been extensively employed because it enables direct observation of CDW gap dynamics and the insulator-to-metal transition [158] [159] [160]. Here, we highlight two representative CDW materials, namely 1$T$-TaS$_2$ and CsV$_3$Sb$_5$.

3.4.1  1$T$-TaS$_2$

1$T$-TaS2 is a model system that exhibits successive CDW phase transitions [161] [162]. As temperature decreases, it undergoes transitions from incommensurate CDW (ICCDW) to nearly commensurate CDW (NCCDW), and finally to commensurate CDW (CCDW) phases at 550 K, 350 K, and 180 K, respectively. In the CCDW phase, the Ta atoms form a Star-of-David cluster. ARPES reveals the evolution of the electronic band structure [163] [164] [165]. Phenomena such as the coexistence of CDW and Mott-insulating phases have driven extensive studies on the role of electron-electron and electron-phonon interactions. Ultrafast spectroscopies offer a powerful alternative to equilibrium methods, enabling the investigation of non-equilibrium states [148], such as a photoinduced metallic phase after strong photoexcitation [166] [167] [168]. TARPES is particularly effective in tracking temporal changes in the band structure, providing insights into transient electronic states [169] [170] [171]. TARPES using high photon energy generated via HHG is ideal for capturing the full Brillouin zone in $1T$-TaS$_2$ [120] [172] [38] [173].

Figures 12(a) and 12(b) show the reconstructed (green lines) and original (black lines) projected Brillouin zone and measured band dispersions along $\bar{\Gamma} - \bar{M} - \bar{K}$ high symmetry path for $1T$-TaS$_2$ [120]. The Mott gap $\Delta_{\text{Mott}}$ and CDW gap $\Delta_{\text{CDW}}$ are clearly observed around the $\bar{\Gamma}$ and $\bar{M}$ points, respectively. Figures 12(c) and 12(d) show the TARPES images around the $\bar{\Gamma}$ and $\bar{M}$ points, respectively, comparing the states before (left) and after (right) the pump excitation. Figures 12(e) and 12(f) show the time-dependent ARPES intensities at selected momenta. While a sub-50 fs melting time of the Mott gap is observed around the $\bar{\Gamma}$ point, a ~225 fs melting time of the CDW gap is observed around the $\bar{M}$ point, accompanied by subsequent oscillations due to the CDW amplitude mode. The melting time of the CDW gap is approximately half the period of the CDW amplitude mode, supporting a structural origin for the insulating signature around the $\bar{M}$ point.

We also conducted TARPES measurements for $1T$-TaS$_2$ [174]. Figure 13(a) shows the TARPES image before pump, and Figure 13(b) shows the differential TARPES image after pump. The photoinduced insulator-to-metal transition, manifested by the band crossing $E_F$ is clearly observed. Figure 13(c) shows the time-dependent photoemission intensities integrated over the energy-momentum regions denoted as I and II in Figs. 13(a) and 13(b). Significant oscillations at a frequency of 2.5 THz are observed, and they have an anti-phase character with respect to each other.

To map this behavior across energy and momentum, we conducted FDARPES analysis. Figures 13(d) and 13(e) show the FDARPES images for amplitude and phase, respectively, at the frequency of 2.5 THz. Two distinctive peak structures near and below $E_F$, shown as black squares and circles in the amplitude part (Fig. 13(d)), are observed. In the phase part (Fig. 13(e)), on the other hand, their phases are different by nearly $\pi$, reflecting the anti-phase character seen in Fig. 13(c). This anti-phase behavior clearly demonstrates that a single band oscillates between the two peak structures, synchronized with the lattice modulation corresponding to the CDW amplitude mode, schematically illustrated in Figs. 12(f) and 12(g). It is also noted that observed oscillation has a cosine-like character, which is consistent with the DECP mechanism [97], while a sine-like character indicates that the coherent phonon is generated by impulsive Raman scatterings [175].

### 3.4.2 CsV$_3$Sb$_5$

The kagome lattice, with its unique geometry, hosts distinct electronic features like flat bands, Dirac bands, and van Hove singularities (VHSs), making it a promising platform for emergent phenomena [176] [177] [178] [179]. In kagome metals AV$_3$Sb$_5$ (A = K, Rb, Cs) [180] [181] [182] [183], a variety of electronic orders—including charge density wave (CDW) [184] [140] [185] [186], superconductivity [187] [188] [189] [190] [191], nematicity [192] [193], and giant anomalous Hall effects [194] [195]— have been observed. Notably, tuning these materials via pressure or doping suppresses CDW and enhances superconductivity, forming two superconducting domes and underscoring a strong CDW-superconductivity interplay [188] [189] [196]. The origin of CDW remains debated. ARPES reveals VHS near $E_F$, supporting an electronically driven CDW [197] [198] [199] [200], while DFT points to lattice instabilities via soft phonon modes [201] [202]. Experimental evidence from STM [203] [204] and x-ray diffraction [205] [206] confirms lattice modulations, yet static measurements cannot fully disentangle electronic and lattice contributions due to strong electron-phonon coupling.

Time-resolved techniques, particularly ultrafast laser excitation, offer a dynamic view by perturbing the CDW and tracking its recovery. Recent studies on CsV$_3$Sb$_5$ have revealed coherent phonon modes at 1.3 THz and

3.1 THz, highlighting electron-phonon interactions [202] [207]. However, their momentum and energy dependencies remain unclear. TARPES enables the direct observation of band dynamics and mode-specific coupling, with frequency-domain ARPES providing further band-resolved insights [208].

Figures 14(a) and 14(b) show the TARPES images before and after photoexcitation. To enhance visibility, the differential image is shown in Fig. 14(c). Near $E_F$, the intensity change is primarily due to the significantly elevated electronic temperature resulting from the absorption of the pump. Notably, in addition to this electronic temperature effect, there are remarkable energy shifts in the electronic bands. A clear increase in intensity between $E_F$ and the saddle band at the M point indicates that the VHS shifts upward toward $E_F$. Figure 14(d) shows the temporal photoemission intensity change around the G, M and K areas, integrated near $E_F$ (from -0.025 to +0.025 eV) over the momentum ranges shown by the solid red lines in Fig. 14(e). Along with the background dynamics, significant oscillations are observed at a frequency of 1.3 THz. To obtain the energy-momentum distribution of the coherent oscillation caused by the 1.3 THz phonon mode, we further mapped FDARPES spectra at 1.3 THz, shown in Figs. 14(f) and 14(g) for amplitude and phase, respectively. These results indicate that the 1.3 THz coherent phonon is strongly coupled to both the Sb-$5p_z$ orbital-derived and V-$3d$ orbital-derived bands, highlighting the three-dimensional nature of the CDW order in $CsV_3Sb_5$.

## 4 Summary and Outlook

In this review, we have briefly summarized recent studies using HHG laser TARPES. We specifically focused on electron-phonon interactions and highlighted various nonequilibrium phenomena that be inaccessible via equilibrium measurements. As this review covers four different classes of materials, we summarize their characteristic time scales, phonon modes, and corresponding frequencies in Table 1. To obtain more detailed information and explore more exotic phenomena, further advancements in both excitation and detection techniques are essential. In terms of excitation, further extension of the wavelength to the mid- and far-infrared regions, and generating stronger fields represent promising directions. A recent report on the use of strong terahertz excitation in TARPES measurements is particularly impressive, making a significant advancement in the ultrafast science community [209] [210]. The development of photoemission electron microscopy is also powerful for observing sub-micrometer length scales, such as semiconductor interfaces or moiré superlattices [211]. Improving time resolution in the attosecond regime could enable studies of more fundamental physics, such as the dynamical transformation of the wave function [212].

Table 1 Summary of the materials and time scales or phonon modes with frequency.

| Material Classes | Specific Materials | Time scale or Phonon Mode (Frequency) |
| --- | --- | --- |
| Graphene | Monolayer graphene | Fast (< 800 fs), slow (> 800 fs) |
|  | Quai-crystalline twisted bilayer graphene | NA |
| Iron-based superconductor | BaFe$_2$As$_2$ | $A_{1g}$ mode (5.5 THz) |
|  | FeSe | $A_{1g}$ mode (5.3 THz) |
| Exitonic Insulator | 1$T$-TiSe$_2$ | NA |
|  | Ta$_2$NiSe$_5$ | $A_{1g}$ mode (1, 2, 3, 3.75, 4 THz) |
| Charge density wave | 1$T$-TaS$_2$ | CDW amplitude mode (2.5 THz) |
|  | CsV$_3$Sb$_5$ | $A_{1g}$ mode (1.3 THz) |

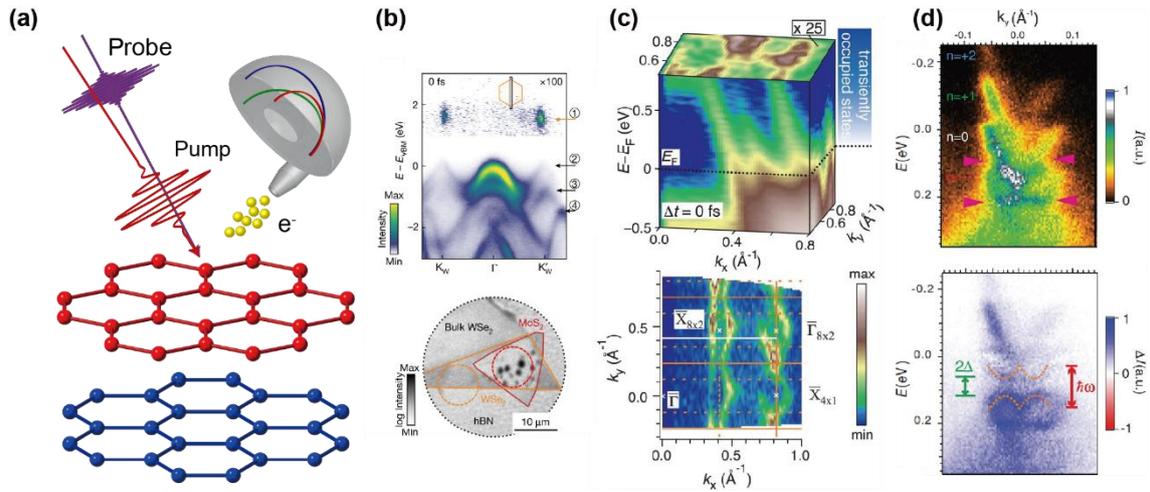

**Fig. 1** (a) Schematic of time- and angle-resolved photoemission spectroscopy (TARPES). (b) TARPES (top) and microscopy (bottom) images for inter- and intralayer excitons in $WSe_2/MoS_2$ [12]. (c) TARPES image of the photoinduced insulator-to-metal transition of atomic indium nanowires on the (111) surface of silicon [17]. (d) TARPES image for the Floquet-Bloch states on the surface of $Bi_2Se_3$ [19]. (b) Reprinted with permission from Ref. [12]. Copyright 2022 Springer Nature Limited. (c) From Ref. [17]. Reprinted with permission from AAAS. (d) From Ref. [19]. Reprinted with permission from AAAS.

**Fig. 2** Schematic illustration of the HHG laser TARPES setup [40]. Amp. and Diff. Pumping refer to Amplifier and Differential Pumping, respectively. Reproduced from [40], with the permission of AIP Publishing.

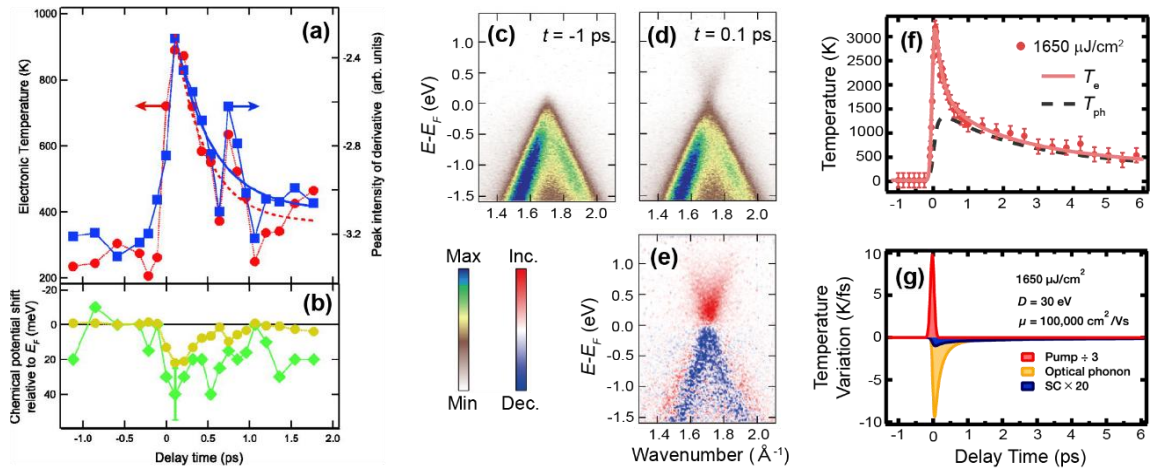

**Fig. 3** (a) Temporal evolution of the electronic temperature (red circles) and intensity derivative at $E_F$ (blue squares) for epitaxial graphene on a SiC substrate. Dashed and solid curves are exponential curve fits. (b) Chemical potential shift corresponding to the electronic temperature at each delay time shown in (a) (yellow circles). Green diamonds indicate the experimentally obtained chemical potential shift [50]. (c), (d) TARPES images before and after photoexcitation for graphene grown on a SiC(0001) C-terminated surface. (e) Differential image by subtracting (c) from (d). (f) Comparison of the electronic temperature relaxation between experiment (red markers) and simulation (black dashed line). (g) Influence of the supercollision process on the cooling of the photoexcited carriers, assuming a deformation potential of 30 eV [56]. (a), (b) Reproduced from [50], with the permission of AIP Publishing. (c)-(g) Reprinted with permission from Ref. [56]. Copyright 2017 by The American Physical Society.

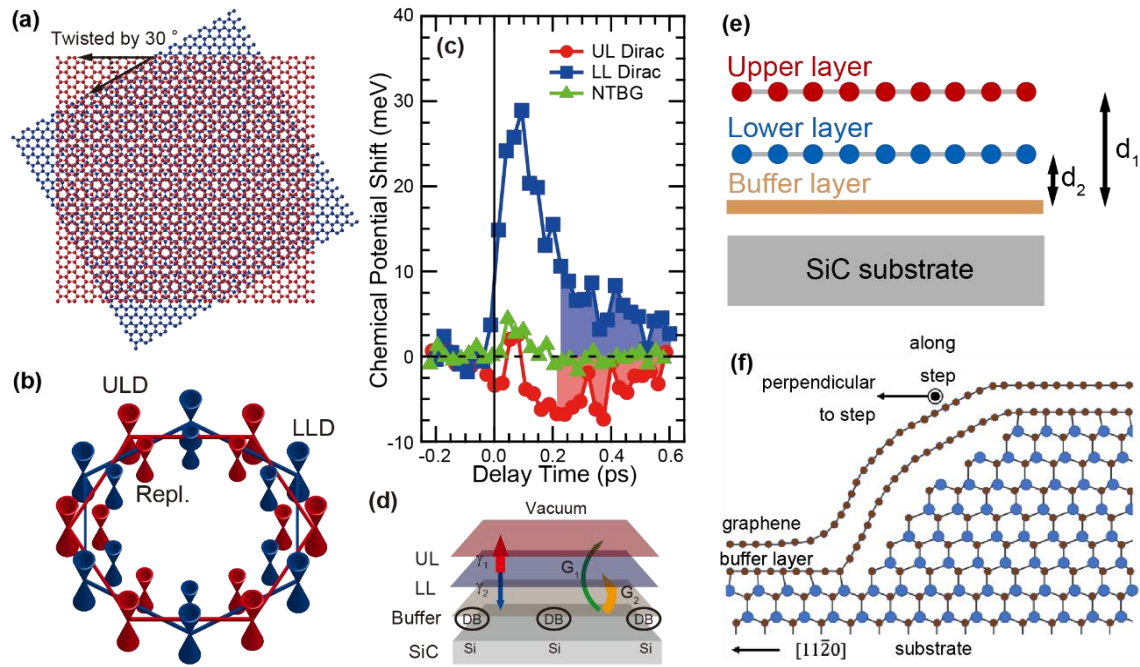

**Fig. 4** (a), (b) Schematic illustrations of crystal structure and band dispersions for a quasicrystalline 30° twisted bilayer graphene (QCTBG). ULD and LLD stands for upper layer Dirac cone and lower layer Dirac cone, respectively. (c) Time-dependent chemical potential shifts for ULD, LLD, and non-twisted bilayer graphene (NTBG). (d) Schematic illustration of the spatial relationships among the upper layer (UL), lower layer (LL), buffer layer, and SiC substrate in QCTBG. Schematic illustration of carrier transport among layers in QCTBG are shown by arrows [66]. (e) Side views of structure models of QCTBG on a SiC(0001) substrate. $d_1$ and $d_2$ are the distances from the SiC substrate to the upper and lower layers, respectively. (f) Cross-sectional structure model of graphene on a stepped SiC substrate, where blue circles represent Si atoms and brown circles represent carbon atoms [67]. (a)-(d) Reprinted with permission from Ref. [66]. Copyright 2019 The American Chemical Society. (e), (f) Reprinted with permission from Ref. [67]. Copyright 2022 by The American Physical Society.

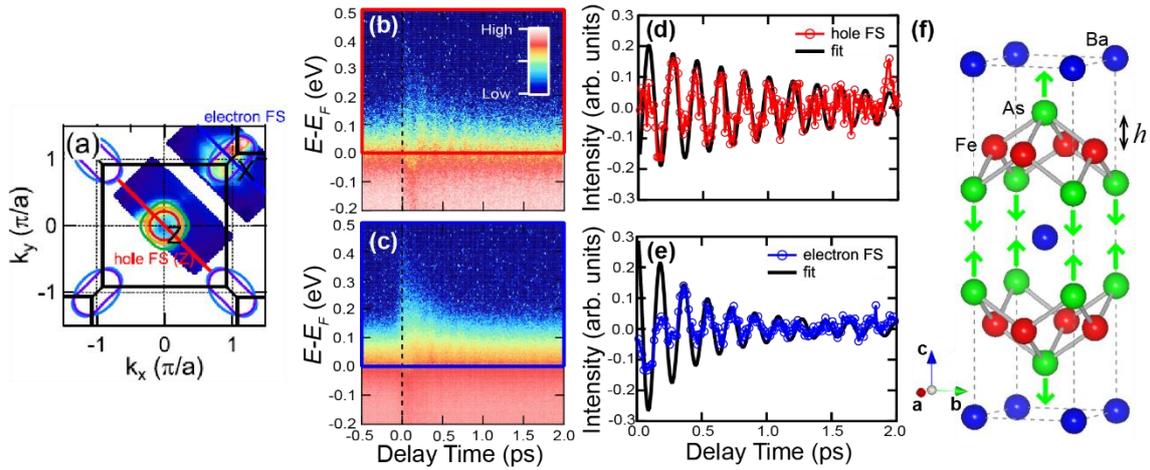

**Fig. 5.** (a) FS mapping of BaFe$_2$As$_2$. The momentum regions integrated for the TARPES spectra across the hole and electron FSs, shown in (b) and (c), are indicated by red and blue lines, respectively. (b), (c) TARPES intensity images with respect to energy relative to $E_F$ and pump-probe delay: (b) around the zone center and (c) around the zone corner. (d), (e) Oscillatory components of the time-dependent photoemission intensities integrated over the red and blue boxes in (b) and (c), respectively. The black solid lines show the fitting results using damped cosine functions. (f) Crystal structure of BaFe$_2$As$_2$ and the definition of the pnictogen height $h$. Thick arrows indicate the displacement of the As atoms corresponding to the $A_{1g}$ phonon [91]. Reprinted with permission from Ref. [91]. Copyright 2018 by The American Physical Society.

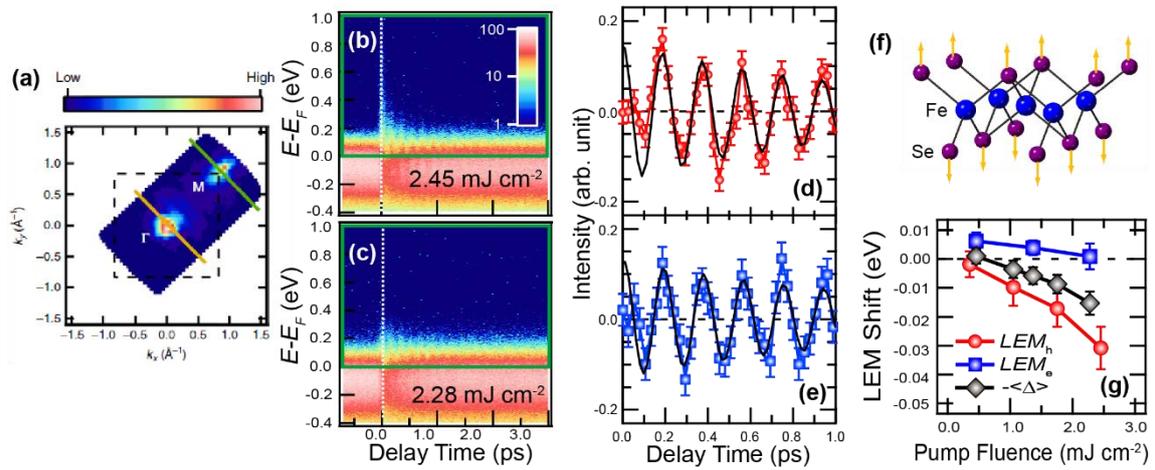

**Fig. 6.** (a) FS mapping of FeSe. The integrated momentum region for the TARPES spectra across the hole and electron FSs shown in (b) and (c) are indicated by orange and green lines. (b), (c) TARPES intensity images with respect to energy relative to $E_F$ and pump-probe delay (b) around the zone center and (c) the zone corner. (d), (e) Oscillatory components of the time-dependent intensities integrated in the red and blue boxes in (b) and (c). Fitting results with the damped cosine functions are shown in black solid lines. (f) Crystal structure of FeSe. Thick arrows indicate the displacement of the Se atoms corresponding to the $A_{1g}$ phonon. (g) Shifts of the leading-edge midpoint (LEM) as a function of pump fluence for the hole and electron bands. The averaged superconducting gap, $\langle\Delta\rangle$, is shown as black solid lines and markers [92]. Reprinted from Ref. [92] licensed under CC BY 4.0.

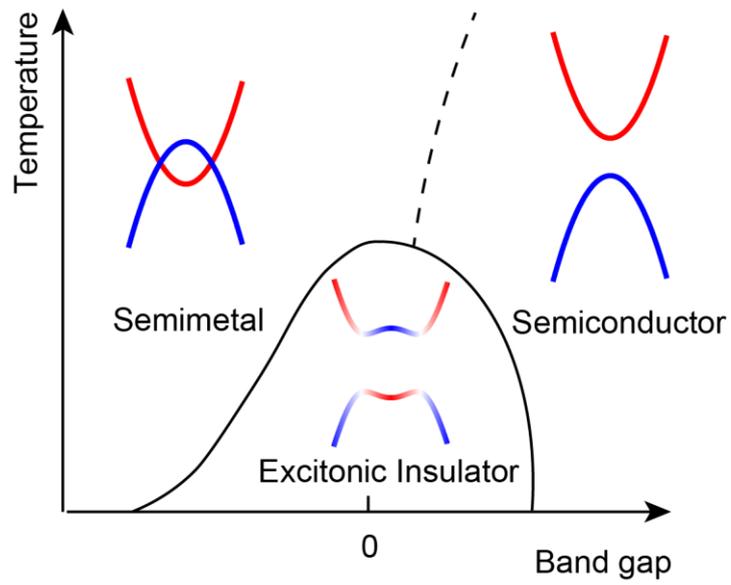

**Fig. 7.** Schematic illustration of the phase diagram for an excitonic insulator as a function of temperature and band gap.

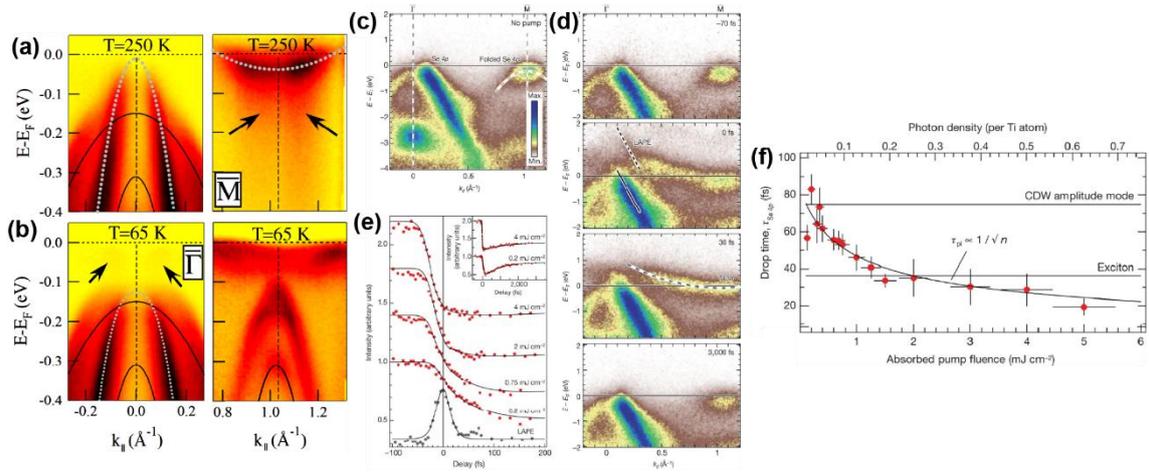

**Fig. 8** (a), (b) Static ARPES images of 1$T$-TiSe$_2$ below and above the CDW transition temperature [115]. (c) ARPES intensity map of 1$T$-TiSe$_2$ recorded with high-harmonic XUV pulses in the CDW phase. (d) TARPES images at increasing pump-probe delays. (e) Photoemission transients of the folded Se 4p band for different excitation fluences [119]. (a), (b) Reprinted with permission from Ref. [115]. Copyright 2007 by The American Physical Society. (c)-(f) Reprinted with permission from Ref. [119]. Copyright 2011 Springer Nature Limited.

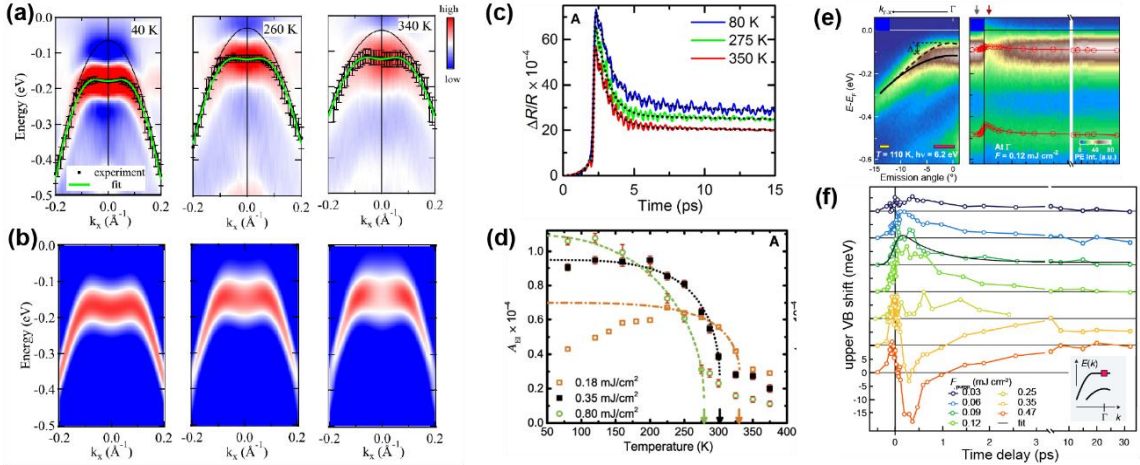

**Fig. 9** (a) Band dispersions of Ta$_2$NiSe$_5$ obtained from ARPES spectra at 40, 260, and 340 K. Dots with error bars indicate band positions determined from the ARPES data. The thick solid curve shows the fit to the excitonic insulator band dispersion, while the thin solid curve indicates the bare valence band dispersion without excitonic coupling. (b) Single-particle excitation spectra at 40, 270, and 340 K obtained by variational cluster approximation (VCA) calculations [125]. (c) Time trace of photoinduced reflectivity changes at different temperatures. (d) Amplitude of the coupled mode at 1 THz as a function of temperature for different excitation densities. The dotted and dashed lines represent fits based on a mean-field-like order parameter fitted to low-temperature data [128]. (e) ARPES spectrum of the occupied electronic structure of Ta$_2$NiSe$_5$ around the Γ point at 110 K and TARPES intensity images with respect to energy and pump-probe delay. (f) Shift of the upper valence band at Γ as a function of pump-probe time delay for different excitation densities [131]. (a), (b) Reprinted with permission from Ref. [125]. Copyright 2007 by The American Physical Society. (c), (d) From Ref. [128]. Reprinted with permission from AAAS. (e), (f) Reprinted from Ref. [131] licensed under CC BY 4.0.

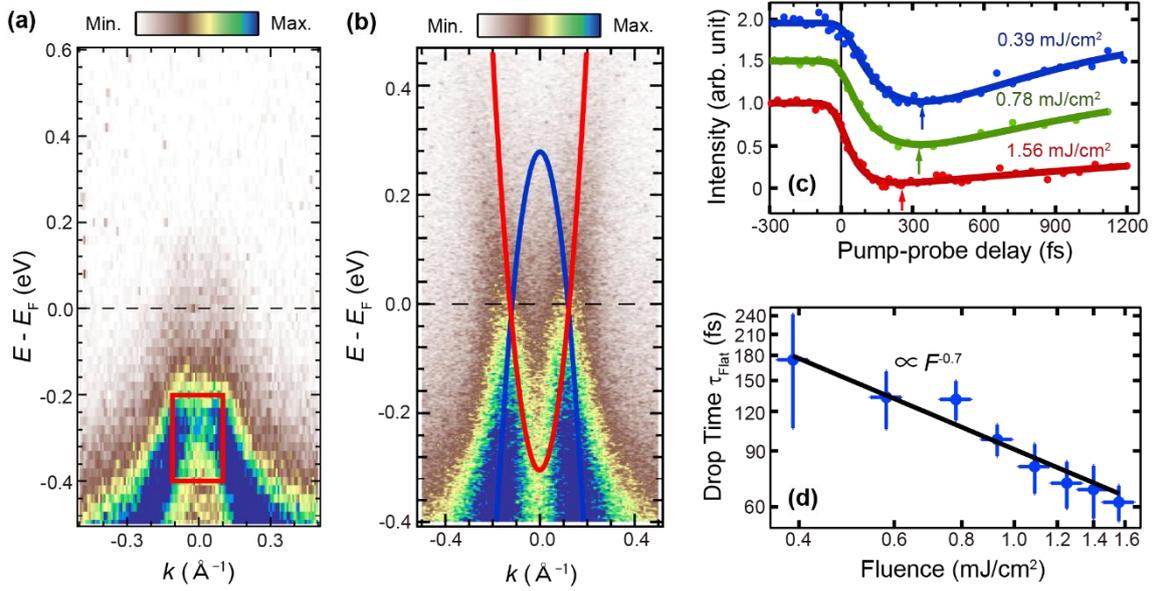

**Fig. 10** (a), (b) TARPES images of $Ta_2NiSe_5$ before and after strong photoexcitation with a fluence of 1.56 mJ/cm$^2$. The red and blue parabolas in (b) indicate the electron and hole bands crossing $E_F$ in the nonequilibrium metallic state. (c) Temporal evolution of the integrated TARPES intensity within the red square shown in (a) for different pump fluences. Arrows indicate the minimum values of the spectral weight. (d) Extracted drop time of the flat band $\tau_{Flat}$ as a function of pump fluence [132]. Reprinted from Ref. [132] licensed under CC BY 4.0.

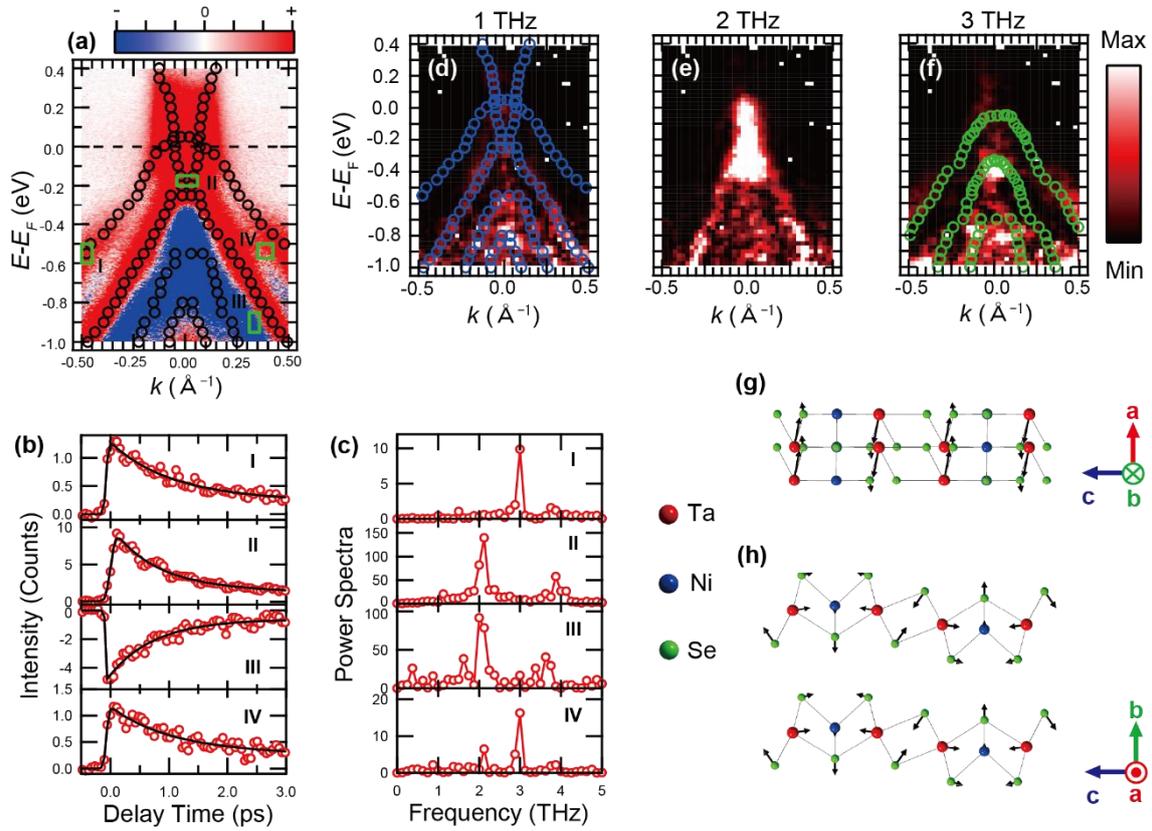

**Fig. 11** (a) Differential TARPES image of Ta$_2$NiSe$_5$, obtained by subtracting the before pump image from the after pump image. (b) Time-dependent TARPES intensities at various energy and momentum regions. Regions I-IV correspond to the areas indicated in (a). Experimental data are shown as red circles, whereas the black solid lines represent the fitting results by double-exponential decay functions convoluted with a Gaussian function. (c) Fourier-transform intensities of the oscillatory components from (b)I–(b)IV, obtained by subtracting the fitting curves from the data. (d)–(f) FDARPES maps showing the energy-momentum distributions of the Fourier-transform intensities of the oscillatory components. The peak positions in the TARPES maps after and before photoexcitation are plotted as blue and green circles in (d) and (f), respectively. (g), (h) Calculated 2-THz and 3-THz phonon modes [133]. Reprinted with permission from Ref. [133]. Copyright 2021 by The American Physical Society.

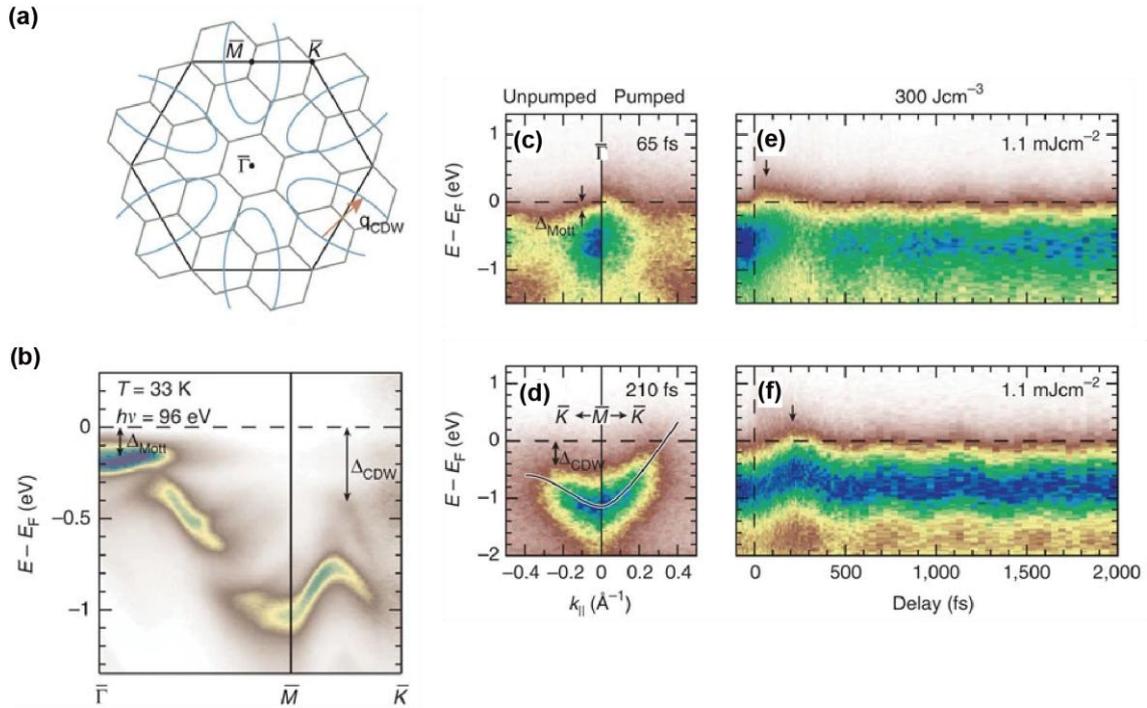

**Fig. 12** (a) Reconstructed (green lines) and original (black lines) projected Brillouin zones of $1T$-TaS$_2$. Unreconstructed schematic FSs (circle and ellipses), wavevectors of the CDW (arrows) and high-symmetry points of the original Brillouin zones (dots) are indicated. (b) Band dispersion along the $\bar{\Gamma} - \bar{M} - \bar{K}$ high-symmetry path. (c), (d) ARPES snapshots taken before (left) optical pumping and at characteristic pump-probe delays (right) around the $\bar{\Gamma}$ and $\bar{M}$ points, respectively. (e), (f) Time-dependent ARPES intensities at the $\bar{\Gamma}$ and $\bar{M}$ points, respectively [120]. (a)-(f) Reprinted with permission from Ref. [120]. Copyright 2012 Springer Nature Limited.

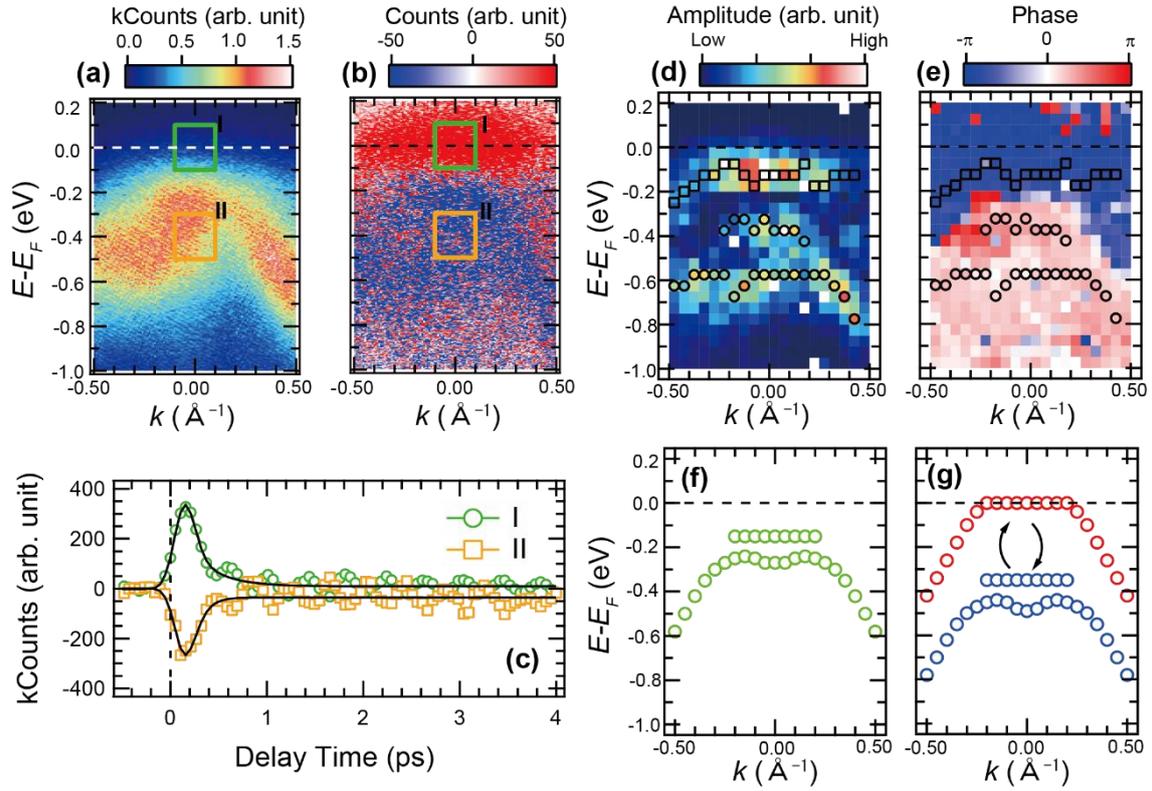

**Fig. 13** (a) TARPES image before the arrival of the pump pulse. (b) Differential TARPES image at a delay time of 240 fs. (c) Time-dependent photoemission intensity integrated over the energy-momentum region denoted as I and II in (a) and (b). (d), (e) FDARPES images for (d) amplitude and (e) phase at the frequency of 2.5 THz. Black markers show the peak positions for the amplitude in (d). (f), (g) Schematic illustration of the band dispersions (f) at equilibrium and (g) after photoexcitation [174]. The uncertainties are within marker sizes. Reproduced from Ref. [174], with the permission of AIP Publishing.

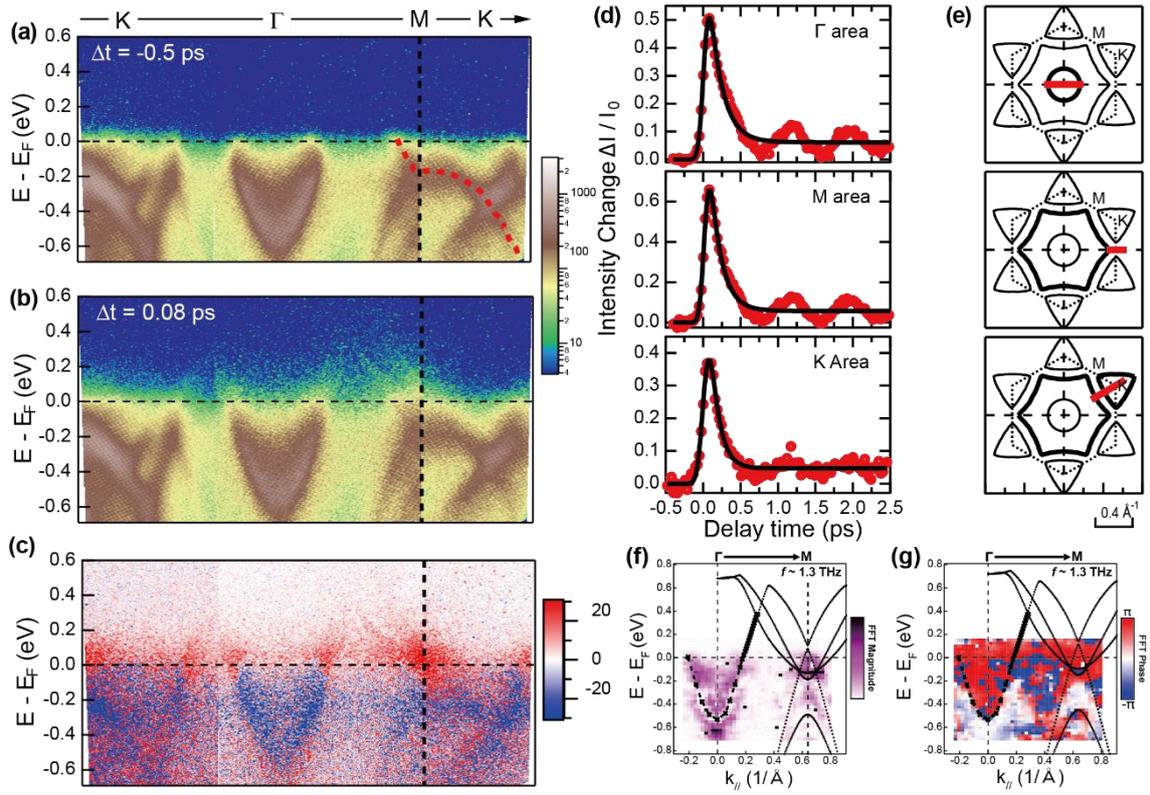

**Fig. 14** (a) TARPES image along the K-Γ-M-K cut using HHG laser TARPES before the arrival of pump pulse. (b) TARPES image at 0.08 ps after the pump. (c) Differential TARPES image obtained from (a) and (b). (d) Dynamics of the intensity changes from ($E_F$ − 0.025 eV) to ($E_F$ + 0.125 eV) around the Γ, M, and K areas. Solids lines are exponential decay fits. (e) Integrated momentum ranges corresponding to each area in (d) are indicated by the red lines. (f), (g) FDARPES maps at 1.3 THz for (f) amplitude and (g) phase [208]. Reprinted from Ref. [208] licensed under CC BY 4.0.

**Acknowledgements**

The authors gratefully acknowledge the late Prof. Shin, who co-authored and made significant contributions to all of the works by our group introduced in this review. His insights and invaluable support greatly advanced our understanding of the subject. This review is dedicated to his memory. Our works introduced in this review were supported by Grants-in-Aid for Scientific Research (KAKENHI) (Grants No. JP19H01818, No. JP19H00659, No. JP19H00651, No. JP24K01375, No. JP24K00565, and No. JP24KF0021) from the Japan Society for the Promotion of Science (JSPS), by JSPS KAKENHI on Innovative Areas "Quantum Liquid Crystals" (Grant No. JP19H05826), and the Quantum Leap Flagship Program (Q-LEAP) (Grant No. JPMXS0118068681) from the Ministry of Education, Culture, Sports, Science, and Technology (MEXT). T.S. acknowledges the research grants from the Izumi Science and Technology Foundation, Itoh Science Foundation, Toyota Riken Scholar, and Iketani Science and Technology Foundation.

**Declaration of competing interest**

The authors declare that they have no known competing financial interests or personal relationship that could have appeared to influence the work reported in this paper.